\documentclass{egpubl_arxiv}
\usepackage{egsr19}
 
 \SpecialIssuePaper         % uncomment for final version of Computer Graphics Forum, special issue
\usepackage[T1]{fontenc}
\usepackage{dfadobe}

\usepackage{url}
\usepackage{xcolor}
\usepackage{fp}
\usepackage{pgfmath}

\usepackage{lscape}
\usepackage{floatflt}
\usepackage{wrapfig}

\usepackage{amsmath}
\usepackage{amssymb}
\usepackage{amsfonts}
\usepackage{upgreek}

\usepackage{microtype}
\usepackage{units}
\usepackage{fancyvrb}
\usepackage{ragged2e}
\usepackage{mdframed}
\usepackage{multirow}
\usepackage{algorithm}
\usepackage{algorithmicx}
\usepackage{algpseudocode}
\usepackage{units}
\usepackage{booktabs}
\usepackage{listings}
\usepackage{verbatim}
\DeclareMathAlphabet{\mathcal}{T1}{pzc}{mb}{it}

\definecolor{superred}{rgb}{1,0,0}
\definecolor{codegrey}{rgb}{0.95,0.95,0.95}
\definecolor{colKeys}{rgb}{0,0.2,0.4}
\definecolor{colKeys2}{rgb}{0,0.8,0}
\definecolor{colIdentifier}{rgb}{0,0,0}
\definecolor{colComments}{rgb}{0,0.6,0}
\definecolor{colString}{rgb}{0.6,0.1,0.1}

\def\mysection#1#2{\section{#1}\label{sec:#2}}
\def\mysubsection#1#2{\subsection{#1}\label{sec:#2}}

\def\negspace{\vspace{-6pt}}
\def\negspaceB{\vspace{-15pt}}
\def\myfigure#1#2{\begin{figure}[thp]\centering\includegraphics*[width = \linewidth]{img/#1}\negspace{}\caption{#2}\negspaceB\label{fig:#1}\end{figure}}
\def\myfigureDouble#1#2{\begin{figure*}[t]\centering\includegraphics*[clip, width = \linewidth]{img/#1}\negspace{}\caption{#2}\negspaceB\label{fig:#1}\end{figure*}}
\def\myfigureCustom#1#2#3#4{\begin{figure}[#2]\centering\includegraphics*[width = #3\linewidth]{img/#1}\negspace{}\caption{#4}\negspaceB\label{fig:#1}\end{figure}}
\def\myfigureCustomDouble#1#2#3#4{\begin{figure*}[#2]\centering\includegraphics*[clip, width = #3\linewidth]{img/#1}\negspace{}\caption{#4}\negspaceB\label{fig:#1}\end{figure*}}

\def\Par#1{\paragraph*{#1.}}

\def\refFig#1{Figure~\ref{fig:#1}}

\def\refSec#1{Section~\ref{sec:#1}}

\def\refEq#1{Equation~\ref{eq:#1}}

\newcounter{myEnumCounter}

\definecolor{todocolor}{RGB}{180,50,30}
\definecolor{commentColor}{RGB}{50,120,20}
\definecolor{darkyellow}{RGB}{160,140,10}
\definecolor{darkgreen}{rgb}{0.05,0.4,0.05}
\definecolor{orange}{RGB}{200,100,0}

\newcommand{\IGNORE}[1]{}

\def\wrt{w.\,r.\,t.\,}
\def\eg{e.\,g., }
\def\ie{i.\,e., }

\def\cf{cf.\ }
\def\citeetal#1{et~al.~\shortcite{#1}}

\def\mathsf#1{\textsf{#1}}
\newcommand{\myvec}[1]{\mathbf{#1}}

\def\dd#1{~\mathrm{d}#1}

\newcommand{\argmin}{\operatornamewithlimits{arg\,min}}

\DeclareMathOperator{\brdf}{\rho}

\newcommand{\mySub}[2]{#1_\mathrm{#2}}

\newcommand{\myL}{L}				% incomming radiance
\newcommand{\myLe}{\mySub{L}{e}} % in-scattered volume radiance
\newcommand{\est}[1]{\widehat{#1}}
\newcommand{\myE}{E}
\newcommand{\Exp}[1]{\mathrm{E}[#1]}
\newcommand{\Var}[1]{\mathrm{Var}[#1]}
\newcommand{\Cov}[1]{\mathrm{Cov}[#1]}
\newcommand{\model}{\mathcal{M}}
\newcommand{\loss}{\uplambda}
\newcommand{\var}[1]{\upnu_{#1}}

\def\posX{\myvec{x}}
\def\dirX{\myvec{\omega}}
\def\pos#1#2{\posX_\mathrm{#1}^\mathrm{#2}}
\def\dir#1#2{\dirX_\mathrm{#1}^\mathrm{#2}}
\def\posY{\pos{}{\prime}}
\def\dirY{\dir{}{\prime}}

\def\percent#1{#1\,\%}

\def\resolutionX#1#2{{#1}$\times${#2}}

\def\bathroomScene{\textsc{Bathroom}}
\def\whiteroomScene{\textsc{WhiteRoom}}
\def\nightroomScene{\textsc{NightRoom}}
\def\futuroScene{\textsc{MeasureOne}}

\newcommand{\RNum}[1]{\uppercase\expandafter{\romannumeral #1\relax}}

\BibtexOrBiblatex
\electronicVersion
\PrintedOrElectronic
\ifpdf \usepackage[pdftex]{graphicx} \pdfcompresslevel=9
\else \usepackage[dvips]{graphicx} \fi

\usepackage{egweblnk}
\title[Learning Patterns in Sample Distributions for Monte Carlo Variance Reduction]%
      {Learning Patterns in Sample Distributions\\for Monte Carlo Variance Reduction}

\author[O. Elek, M. M. Thomas, A. Forbes]{
\parbox{\textwidth}{\centering
\vspace{-3mm}
Oskar Elek \qquad Manu M. Thomas \qquad Angus Forbes\\
\vspace{2mm}
University of California, Santa Cruz, USA}
}

\begin{document}

%\teaser{
 %\includegraphics[width=\linewidth]{eg_new}
 %\centering
  %\caption{New EG Logo}
%\label{fig:teaser}
%}

\maketitle

%================================================================================================
%================================================================================================

\begin{abstract}
This paper investigates a novel a-posteriori variance reduction approach in Monte Carlo image synthesis.
Unlike most established methods based on lateral filtering in the image space, our proposition is to produce the best possible estimate for each pixel separately, from all the samples drawn for it.
To enable this, we systematically study the per-pixel sample distributions for diverse scene configurations.
Noting that these are too complex to be characterized by standard statistical distributions (e.g. Gaussians), we identify patterns recurring in them and exploit those for training a variance-reduction model based on neural nets.
In result, we obtain numerically better estimates compared to simple averaging of samples.
This method is compatible with existing image-space denoising methods, as the improved estimates of our model can be used for further processing.
We conclude by discussing how the proposed model could in future be extended for fully progressive rendering with constant memory footprint and scene-sensitive output.
%-------------------------------------------------------------------------
%  ACM CCS 1998
%  (see https://www.acm.org/publications/computing-classification-system/1998)
% \begin{classification} % according to https://www.acm.org/publications/computing-classification-system/1998
% \CCScat{Computer Graphics}{I.3.3}{Picture/Image Generation}{Line and curve generation}
% \end{classification}
%-------------------------------------------------------------------------
%  ACM CCS 2012
%   (see https://www.acm.org/publications/class-2012)
%The tool at \url{http://dl.acm.org/ccs.cfm} can be used to generate
% CCS codes.
%Example:


\end{abstract}  

%================================================================================================
%================================================================================================

\mysection{Introduction}{Introduction}

Monte Carlo (MC) integration is the de-facto standard for physically based image synthesis, stemming mainly from its broad applicability~\cite{Christensen2016} and solid mathematical foundations~\cite{Kalos2008}.
On the downside, the stochastic nature of MC methods causes the resulting solution to contain a certain amount of error that fully vanishes only in the theoretical limit case of infinitely many evaluated samples.

This error is the \emph{residual variance} of the used MC estimator, often referred to as ``Monte Carlo noise''.
This nomenclature is somewhat unfortunate, since ``noise''---in the signal processing context---typically refers to an undesired distortion of a signal caused by instruments or otherwise.
In contrast, MC variance is an \emph{inherent phenomenon}, inseparable from, and specific to any MC estimator.

The above distinction goes beyond mere linguistic nitpicking.
Numerous existing approaches~\cite{Zwicker2015} approach the task of MC variance reduction through the image-processing lens, that is, as an image denoising problem.
While that can certainly produce satisfactory results due to the similarities of the two problems, the inherited assumptions are often not applicable in MC image synthesis (as partially pointed out by Boughida and Malik~\shortcite{Boughida2017}, for instance).
As a practical consequence of this mismatch, it is difficult to predict such methods' performance upfront, necessitating trial-and-error application and hand-tuning of parameters.

Consequently, the core mission of this paper is to critically examine the problem of variance reduction in Monte Carlo image synthesis.
We build on the following paradigm (already adopted in this context by several methods~\cite{Zirr2018, Boughida2017, DeCoro2010, Delbracio2014}): \textbf{every image sample produced by an MC estimator carries relevant information about the underlying distribution from which it is drawn.}
These distributions---specific to every \emph{point} in the image---will be denoted as \emph{characteristic distributions}.
Our proposition is to identify the characteristic distributions in the context of MC image synthesis, and use that knowledge to produce better estimates for \emph{every image pixel individually}, rather than simply averaging the samples generated for that pixel.
In that sense, our method belongs to the \emph{a-posteriori} category.
This is in contrast to \emph{a-priori} methods which attempt to derive a variance reduction strategy from theoretical principles; although we do ground our reasoning in theory as well, whenever possible.

Towards the above goal, we undertake three main steps.
\begin{itemize}
	\item We first \textbf{discuss the problem of variance reduction in MC image synthesis}, pointing out the key sources of variance typical for this domain; then review the works tackling the problem (\refSec{Background}).
	Afterwards we critically examine the assumptions commonly used in MC denoising methods (\refSec{PropertiesSampleDistributions}) and propose how to alleviate them.
	\item Next, we present a \textbf{qualitative study of characteristic sample distributions} in several scenes and their configurations (\refSec{CaseStudy}).
	This is our first contribution, intended to identify patterns occurring in these distributions.
	\item Finally, we describe a \textbf{prototype framework for learning these characteristic distributions} (\refSec{VarianceReductionModel}).
	The learned statistical description of the distributions then serves to provide better pixel estimates from MC samples collected during rendering.
	This framework, agnostic to the internal workings of the employed integrator, is our second contribution.
\end{itemize}

The benefit of the proposed method is that it is complementary to the existing image-space methods: our intent is to provide improved estimates for each image pixel, which can subsequently serve as an input to further processing.
In contrast to many existing denoisers, we do not (explicitly or implicitly) assume any specific form or properties of the underlying sample distribution.
The only assumption embedded in our method is that the characteristic distributions \emph{do contain identifiable patterns}, which can be learned by a sufficiently expressive statistical model (based on neural nets, in our case).

%================================================================================================
%================================================================================================

\myfigureCustomDouble{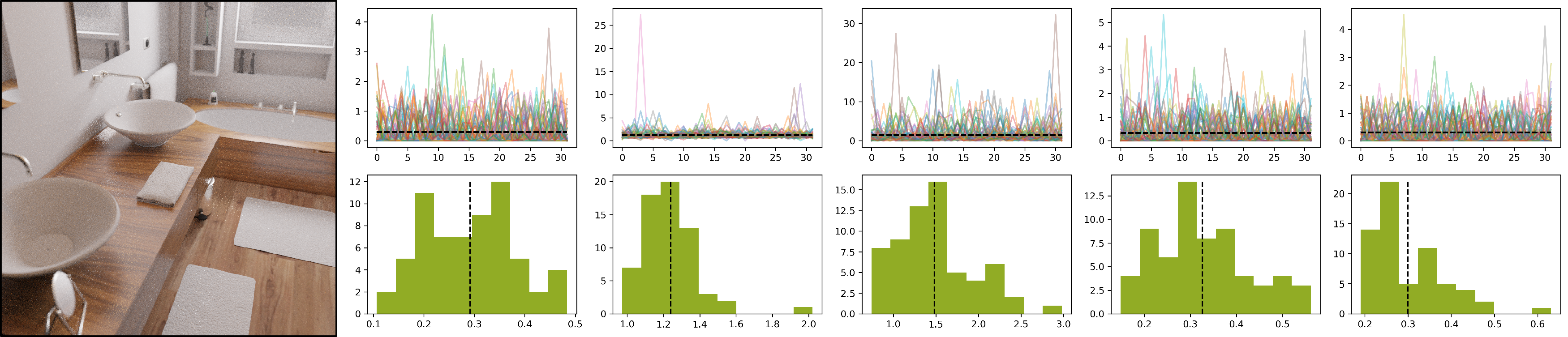}{t}{1}{
Example distributions of 64 superimposed batches of 32 radiance samples $\est{\myL}$ each \emph{(top row)} and the corresponding pixel irradiance estimate $\est{\myE}$ histograms \emph{(bottom row)} for 5 randomly chosen pixels \emph{(columns)} of the \bathroomScene{} scene \emph{(left)}.
The corresponding expected values are marked with dashed black lines.
The apparent long-tailedness of the radiance samples (leading to numerous outliers above the expectncy) is reflected in the slight skew in the aggregate irradiance values.
}

\mysection{Background}{Background}

This section first defines the problem of Monte Carlo variance reduction and provides an overview of works that address it.

\Par{Realistic image synthesis}
The central problem of realistic image synthesis is governed by the rendering equation~\cite{Kajiya1986}, which, for a given position $\posX$ and direction $\dirX$, evaluates the (monochromatic) incident radiance $L$:
\begin{gather}
  \myL(\posX,\dirX) = \myLe(\posY) + \int_{\Omega} \myL(\posY,\dirY) \brdf(\posY,\dirX,\dirY) \dd{\dirY}.
\label{eq:RE}
\end{gather}
Here, $\posY$ is the point on the nearest surface visible from $\posX$ in the direction $\dirX$; $\myLe$ and $\brdf$ are the emitted radiance and the cosine-weighted bi-directional reflectance distribution function (BRDF) at that point, respectively.
The key to solving \refEq{RE} is the hemispherical integral on the right, which again requires solving for incident radiance at $\posY$, leading to the well known self-referential nature of the rendering equation which makes its solution so challenging.

Rendering an image for a given scene entails gathering the radiant energy passing through every pixel $i$.
Since each pixel corresponds to a small solid angle $\Omega_i$, this yields the irradiance $\myE_i$ of that pixel (for a fixed camera position $\pos{c}{}$):
\begin{gather}
	\myE_i = \int_{\Omega_i} \myL(\pos{c}{},\dirX) \dd{\dirX}.
\label{eq:Ei}
\end{gather}
The question this paper is concerned with is: assuming a particular Monte Carlo solver, what is the `best' way to obtain the value of $\myE$ for every pixel $i$ in the rendered image?
For this, we will employ the standard MC path tracing with next event estimation~\cite{Pharr2016}; in general, we will limit our analysis to \emph{unbiased} MC solvers to only focus on variance analysis and reduction.

\Par{Monte Carlo estimation}
Being a stochastic method, path tracing generates samples $X_n, n \in 1 \ldots N$ for each pixel (for simplicity we will assume the same $N$ across the entire image, although our analysis does not depend on this assumption).
These samples $X_n$ are paths originating at the camera position $\pos{c}{}$ and uniformly sampled directions $\dirX_i \in \Omega_i$, probabilistically traced through the scene.
Together they form an estimator for \refEq{Ei}
\begin{gather}
  \est{\myE} = \frac{1}{N} \sum_n^N \est{\myL}(X_n) = \frac{1}{N} \sum_n^N \frac{\myL(X_n)}{p(X_n)},
\label{eq:estEi}
\end{gather}
where $p(X_n)$ is the probability of that path; it is the joint probability of all stochastic decisions made by the path tracer.
Therefore, the estimate for $\myE$ is composed of $N$ evaluations of the estimator $\est{\myL}$ for \refEq{RE}.
Since we assume $\est{\myL}$ to be an unbiased estimator, also the estimator for the pixel value is unbiased, that is, $\Exp{\est{\myE}} = \myE$.

The variance (the error) of the estimator for a particular $\myE_i$ depends on two factors: the variance of the individual estimates $\est{\myL}(X_n)$, and the variation of $\myL$ itself across the region of the pixel $i$.
If, for a moment, we assume the latter to be constant, then
\begin{align}
  \Var{\est{\myE}} &= \Var{~\frac{1}{N} \sum_n^N \est{\myL}(X_n)~} = \frac{1}{N^2} \cdot \Var{~\sum_n^N \est{\myL}(X_n)~} = \notag \\
	&= \frac{1}{N^2} \cdot \left( \sum_n^N \Var{\est{\myL}(X_n)} + \sum_{n_1 \neq n_2}^N \Cov{\est{\myL}(X_{n_1}),\est{\myL}(X_{n_2})} \right) \leq \notag \\
	& \leq \frac{1}{N^2} \cdot \sum_n^N \Var{\est{\myL}(X_n)}, \label{eq:VarE}
\end{align}
where the final inequality results from the fact that sample covariance is zero for samples independently distributed across the pixel (as is the case in this analysis), or negative if correlated (low-discrepancy) sampling is used.
Dropping the temporary assumption of constant $L$ within the pixel $i$, the actual variance of $\est{\myE_i}$ is bound to be slightly higher than \refEq{VarE} indicates (and without caching the estimates to adaptively refine the pixel-space sampling, we cannot do much better than sample it uniformly, since no prior estimate on the distribution of inter-pixel $L$ is available to us).

\Par{Reducing estimator variance}
Since the variance of ${\est{\myE}}$ is the main efficiency bottleneck of MC rendering, diverse strategies exist to reduce it.
For instance, according to the zero-variance sampling theory~\cite{Hoogenboom2008, Kalos2008}, generating the paths $X_n$ exactly according to the energy distribution in the scene (\ie $p(X)\propto\myL(X)$, \cf \refEq{estEi}) would result in an optimal estimator $\est{\myL}$.
On the other hand, any deviations from the optimal sampling probabilities for \refEq{RE} result in accumulation of variance---including:
\begin{itemize}
	\item using loose approximations for sampling the BRDF $\brdf$,
	\item omitting the radiance term $\myL$ when sampling the integrand,
	\item incorrect path termination strategies, and
	\item visibility fluctuation during next event estimation.
\end{itemize}
Numerous works in realistic image synthesis therefore strive to approach the theoretical zero-variance limit by developing advanced importance-sampling strategies, for instance using the path guiding approach~\cite{Vorba2014, Mueller2017, Guo2018}.
Given our \emph{a-posteriori} sample-based approach, we refer the reader to established literature~\cite{Pharr2016} for further information regarding this class of methods.

\Par{Image filtering}
The tradition of image-space (lateral) filtering to denoise images has a long history in image processing.
Perhaps the most influential signal-preserving method---bilateral filtering~\cite{Tomasi1998}---has been extended through non-local neighborhoods~\cite{Buades2005}, cross\,/\,joint filtering~\cite{Xu2005}, to comprehensive multi-faceted methods~\cite{Rousselle2013}, to provide a few examples.

The sophistication of the recent methods nevertheless means that the simplicity of the original bilateral filter is lost, and stricter assumptions are needed (\cf \refSec{Methodology}); assumptions that have the potential to fail.
The inherent noisiness of the feature descriptors, and their unknown correlation with radiance distribution, are just some examples of the necessary guesswork.
It is interesting that many of these methods can be cast as variants of statistical regression~\cite{Bitterli2016}.
Reconstruction methods like that can be powerful, as well as modified for efficiency~\cite{Schied2017}---ultimately though, they still need to assume a certain smoothness of the reconstructed radiance distribution, which is violated by visibility and other forms of discontinuities extremely common in rendering.

\Par{Sample-based methods}
A very insightful progress was made by Delbracio~\citeetal{Delbracio2014}: for lateral filtering to be sensitive to the properties of the reconstructed signal, we need to take into account statistical similarity between sample distributions of the involved image regions.
Prior to that, a density-based classification was proposed for outlier removal~\cite{DeCoro2010}, but only on the binary accept\,/\,reject basis.
As these methods have a significant memory bottleneck, an alternative statistical treatment has been developed by Boughida and Boubekeur~\shortcite{Boughida2017}, however relying heavily on the suboptimal Gaussian representation (see more detailed discussion in \refSec{PropertiesSampleDistributions}); the same shortcoming has the recent robust-statistical estimation of Jung~\cite{Jung2015}.

Their conclusions however hold: the problematic long-tailedness of the sample distributions calls for a robust reweighing approach.
This conclusion has been recently confirmed by Zirr~\citeetal{Zirr2018} for outlier compensation and likewise our paper---in fact, it is the ideas of Jung, Zirr and colleagues that we take forward and develop from a different perspective.
In the following section, we argue for a simple method that combines the advantages of the bottom-up statistical approaches and their data-driven counterparts.

\Par{Learning-based filtering}
Complex filtering operators can be learned from data---even if accurate references are not feasible~\cite{Lehtinen2018}.
This approach has been used for both indirect~\cite{Kalantari2015} and direct~\cite{Bako2017, Vogels2018} description of lateral filters.
In addition to the offline operators above, a low-sample temporally stable method has been demonstrated by Chaitanya~\citeetal{Chaitanya2017}.
Our aim is to combine such data-driven modeling approaches with a principled analysis of sample distributions and---as we explain in \refSec{Methodology}---reduce the variance of pixel estimators individually.

Ultimately, the amount of variance-reduction literature is particularly extensive, and we cannot hope to provide a complete overview.
Interested readers are encouraged the seek the detailed review by Zwicker~\citeetal{Zwicker2015} and the works of Bitterli~\citeetal{Bitterli2016} and Vogels~\citeetal{Vogels2018} for reviews of more recent papers.

%================================================================================================
%================================================================================================

\mysection{Overview of Methodology}{Methodology}

Our exposition has two parts.
We first review review the assumptions commonly made to tackle the MC variance reduction problem (\refSec{PropertiesSampleDistributions}).
Finding them too optimistic, we conduct a study to empirically determine the properties of the characteristic sample distributions in different rendering configurations (\refSec{CaseStudy}).
These constitute empirical priors for the next step.

The information contained in each sample is viewed through the perspective of the scene-specific priors to produce better pixel estimates than simply averaging the available samples.
In the second part (\refSec{VarianceReductionModel}) we present a learning-based approach to facilitate this, and show that it is indeed possible to significantly improve the pixel estimates on such grounds.

%================================================================================================
%================================================================================================

\myfigureCustomDouble{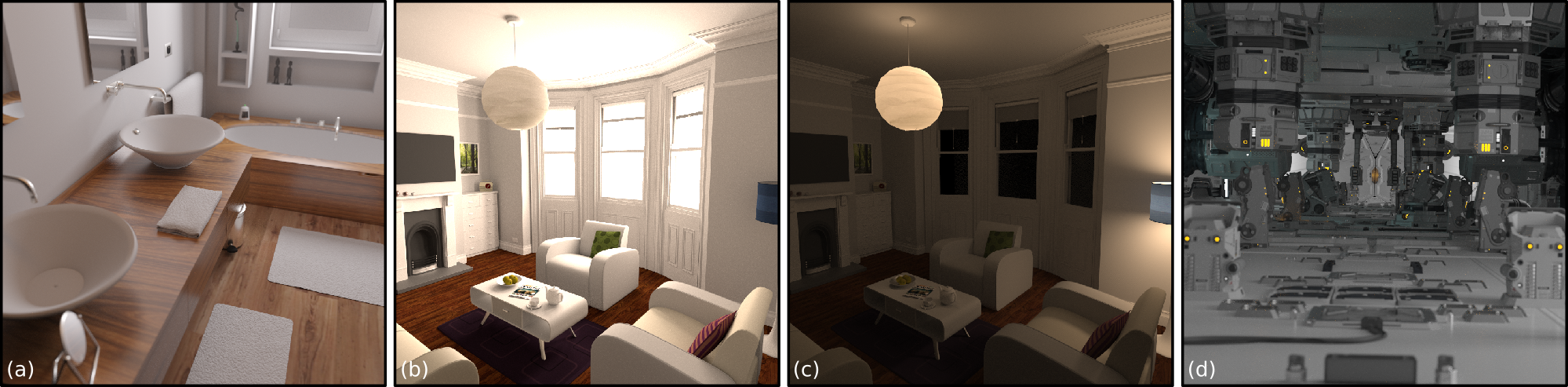}{t}{1}{
The scenes used in our study: \bathroomScene{} \emph{(a)}, \whiteroomScene{} \emph{(b)}, \nightroomScene{} \emph{(c)} and \futuroScene{} \emph{(d)}.
}

\mysection{Properties of Sample Distributions}{PropertiesSampleDistributions}

We aim to gain a better understanding of the characteristic per-pixel sample distributions in MC image synthesis, and set reasonable expectations about the patterns we might encounter in them.
To this end, we start with a high-level discussion of the assumptions most commonly adopted by existing variance-reduction methods, and continue with the detailed study in \refSec{CaseStudy}.

\Par{Spatial uniformity}
One of the two most common assumptions is that the estimator error can be modeled by a spatially invariant additive `noise' $\var{}$, \ie $\est{\myE} = E + \var{E}$.
While suitable in image processing to model, \eg thermal sensor noise, it is an unjustified assumption in MC image synthesis.

First, as a straightforward consequence of \refEq{VarE}, the estimator variance is a function of the incoming radiance and its estimator.
As such, it will differ not only for each image pixel (for $\est{\myE}$), but even vary \emph{within the pixel itself} (for $\est{\myL}$).
This has also been pointed out by Boughida and Boubekeur~\shortcite{Boughida2017} and Kalantari and Sen~\shortcite{Kalantari2013}, among others.

Second, as a consequence of the previous property and the fact that light transport is a relative process (implying that the estimator variance scales with the value of the estimated quantity itself), modeling variance with an additive term is problematic in case of $\var{L}$, and by transition (although to a lesser extent), for $\var{E}$ as well.
This is also supported by empirical data (\cf \refSec{CaseStudy}), mainly due to the characteristic skew and long-tailedness of characteristic sample distributions (see \refFig{sample_distributions_and_means}).

\Par{Gaussian distribution}
While the per-pixel characteristic distributions are known to be complex \cite{DeCoro2010, Delbracio2014, Boughida2017}, their means---the estimator values $\est{\myE}$---are very frequently assumed to be normally distributed.
This seems like a safe bet following from the central limit theorem (CLT), but is seldom verified.

CLT holds for random variables which are \emph{independent and identically distributed with finite variance}.
However, the aggregated random variables $\est{\myL}$ are certainly not identically distributed (see previous point) and might not be independent (when correlated sampling is used).
Furthermore, in spite of the theoretically backed convergence rates (\refEq{VarE}), no bounds on the individual sample variances are available; in fact, under certain circumstances involving source singularities, variance might be provably infinite~\cite{Georgiev2013}.
The existence of strong outliers demonstrated by Zirr~\citeetal{Zirr2018}, as well as here (\refFig{sample_distributions_and_means}), is an additional empirical evidence that not only individual samples but even their averages can follow long-tailed distributions which do not have the well mannered behavior assumed by Gaussian statistics.

\Par{Correlation with `scene features'}
A common variance-reduction approach is the use of feature buffers---auxiliary information like surface positions, normals and albedos---to guide the filtering process (\refSec{Background}).
Again adopted from image processing, this approach is purely heuristical in the context of MC image synthesis: it carries the hidden assumption that variance be correlated with these features, which is trivially not true for numerous physical phenomena (shadows, caustics, transparent objects, volumetric media).
Some of these pathological cases have been demonstrated by Boughida and Boubekeur~\shortcite{Boughida2017}.

Again, this does not mean that a variance-reduction approach based on feature buffers cannot work; we simply argue that by relying on feature buffers, an unknown amount of variance is either left untreated, or actual image features get filtered out.
In general, the complexity of the path space is virtually unbounded (due to its recursive nature), and thus it seems unlikely that any reasonable number of hand-picked features can be found to sufficiently describe that complexity.

\Par{Regularity of the radiance distribution}
Finally, even if no assumptions are made about the variance distribution, methods based either on lateral filtering or statistical regression assume some degree of smoothness in the radiance distribution.
This assumption can hold true for large portions of the image, but is violated by visibility and illumination discontinuities---an issue that becomes more abundant as the complexity of the scene in question increases.

Concluding this discussion, we have argued that many of the assumptions frequently adopted for solving the MC variance reduction problem are fragile.
We therefore propose to take a step back and---to complement the existing theoretical analysis~\cite{Jung2015, Boughida2017, Zirr2018}---examine the characteristic distributions empirically.

%================================================================================================
%================================================================================================

\def\studySize{0.97}

\myfigureCustomDouble{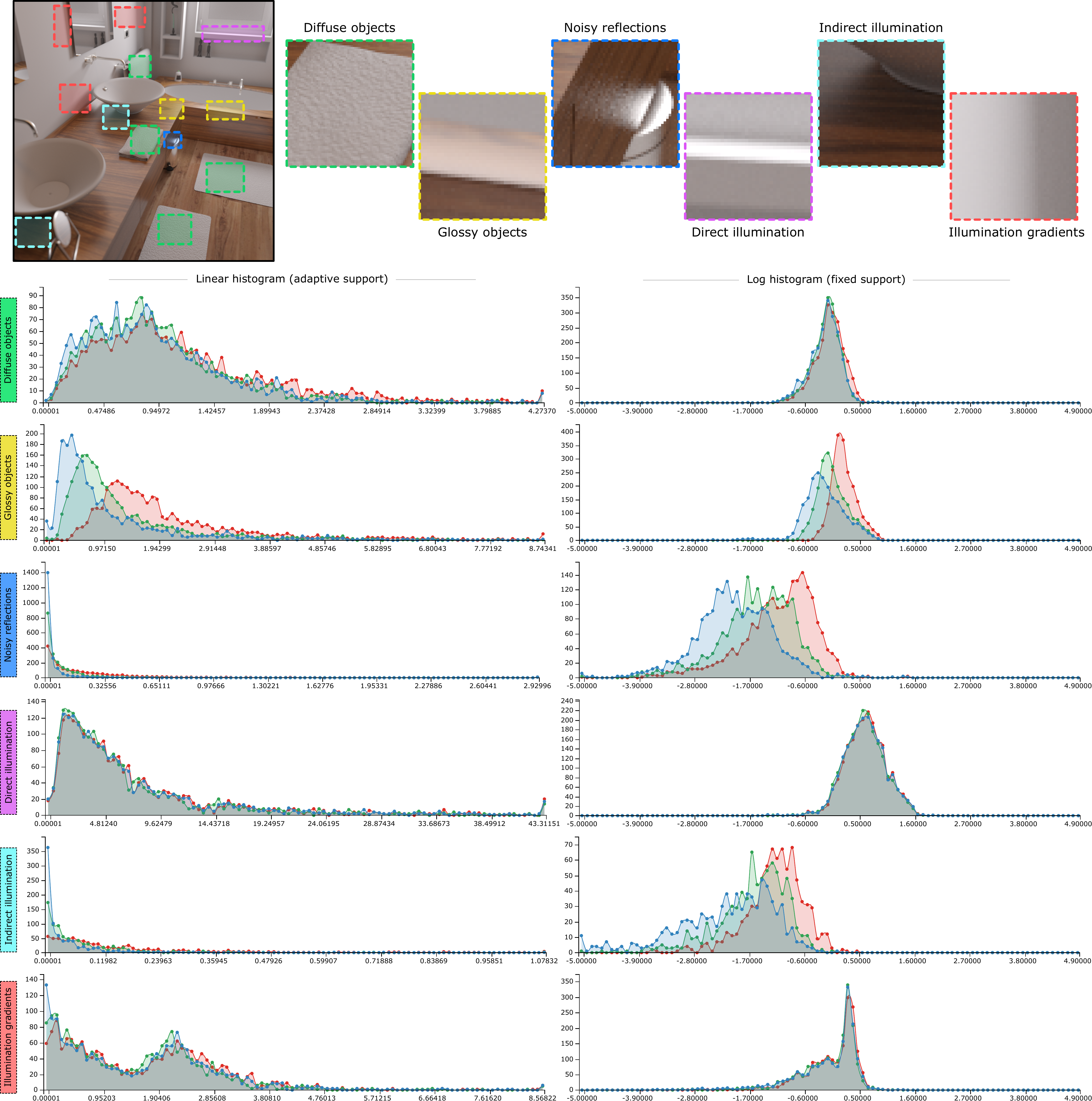}{t}{\studySize}{
Study for the \bathroomScene{} scene, analyzed in \protect{\refSec{StudySceneA}}.
The regions of interest are selected using different colors, which then mark the corresponding magnified insets as well as the characteristic sample distributions below; we use this scheme across the whole study.
}

\myfigureCustomDouble{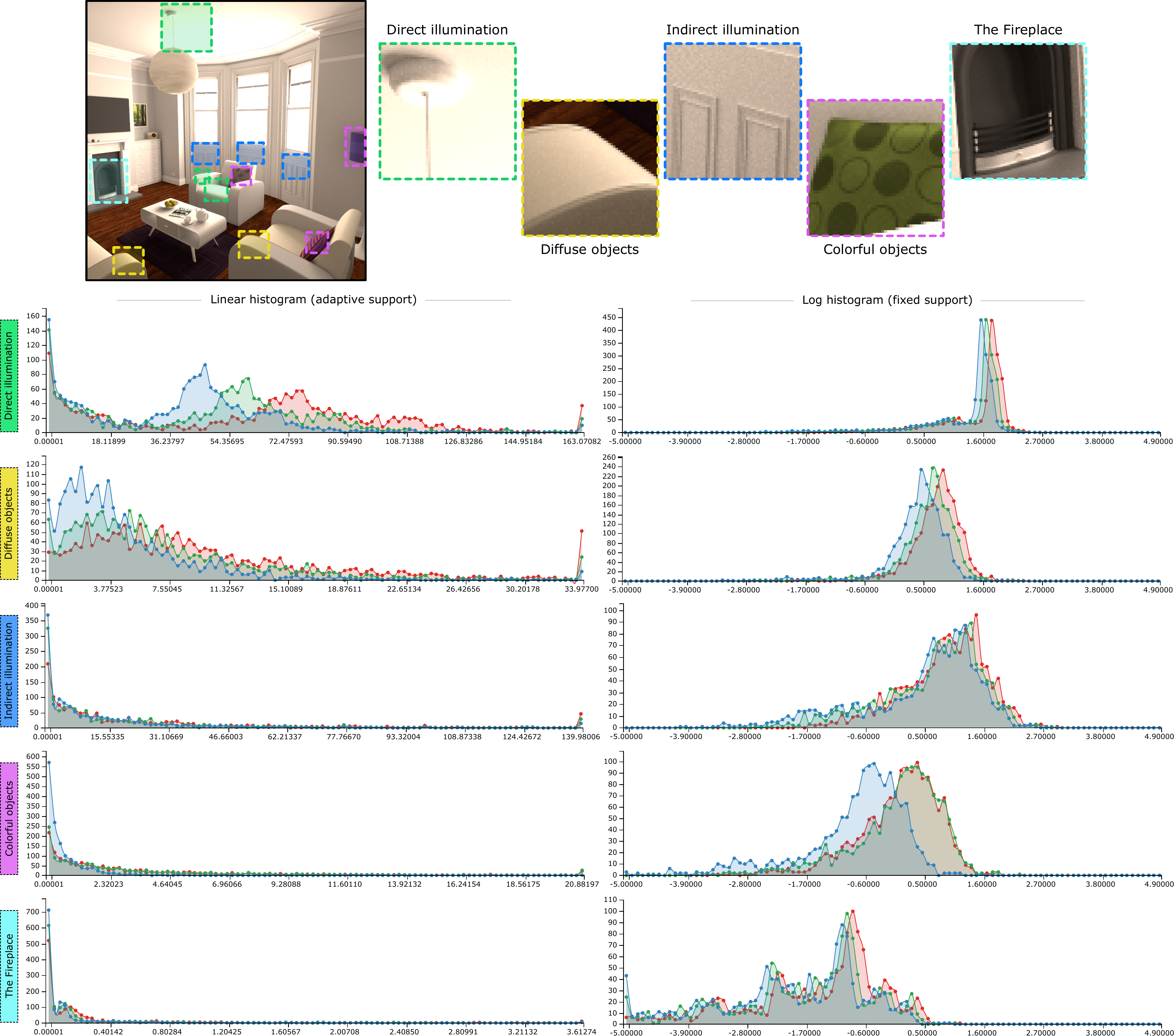}{t}{\studySize}{
Study for the \whiteroomScene{} scene, analyzed in \protect{\refSec{StudySceneB}}.
}

\myfigureCustomDouble{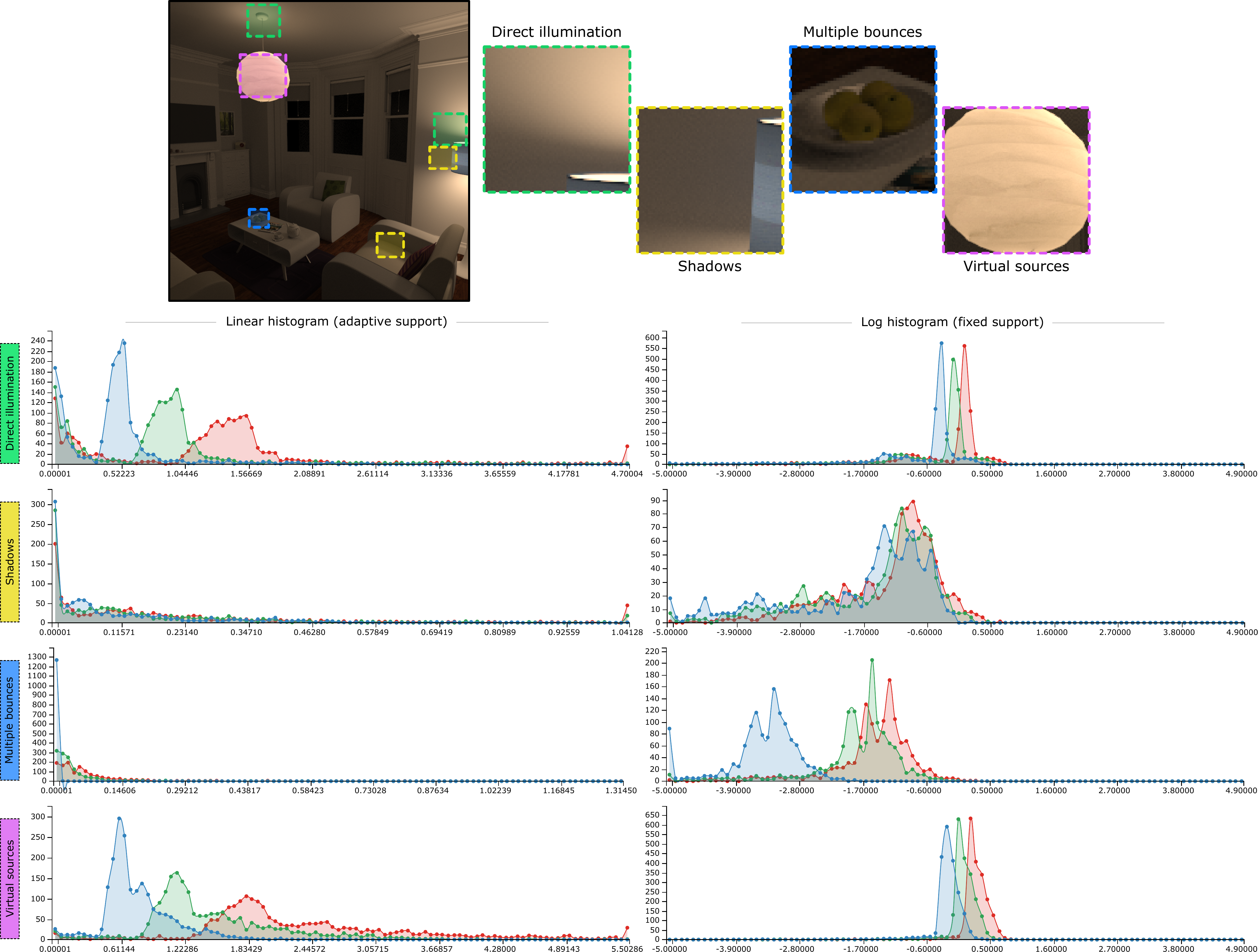}{t}{\studySize}{
Study for the \nightroomScene{} scene, analyzed in \protect{\refSec{StudySceneC}}.
}

\myfigureCustomDouble{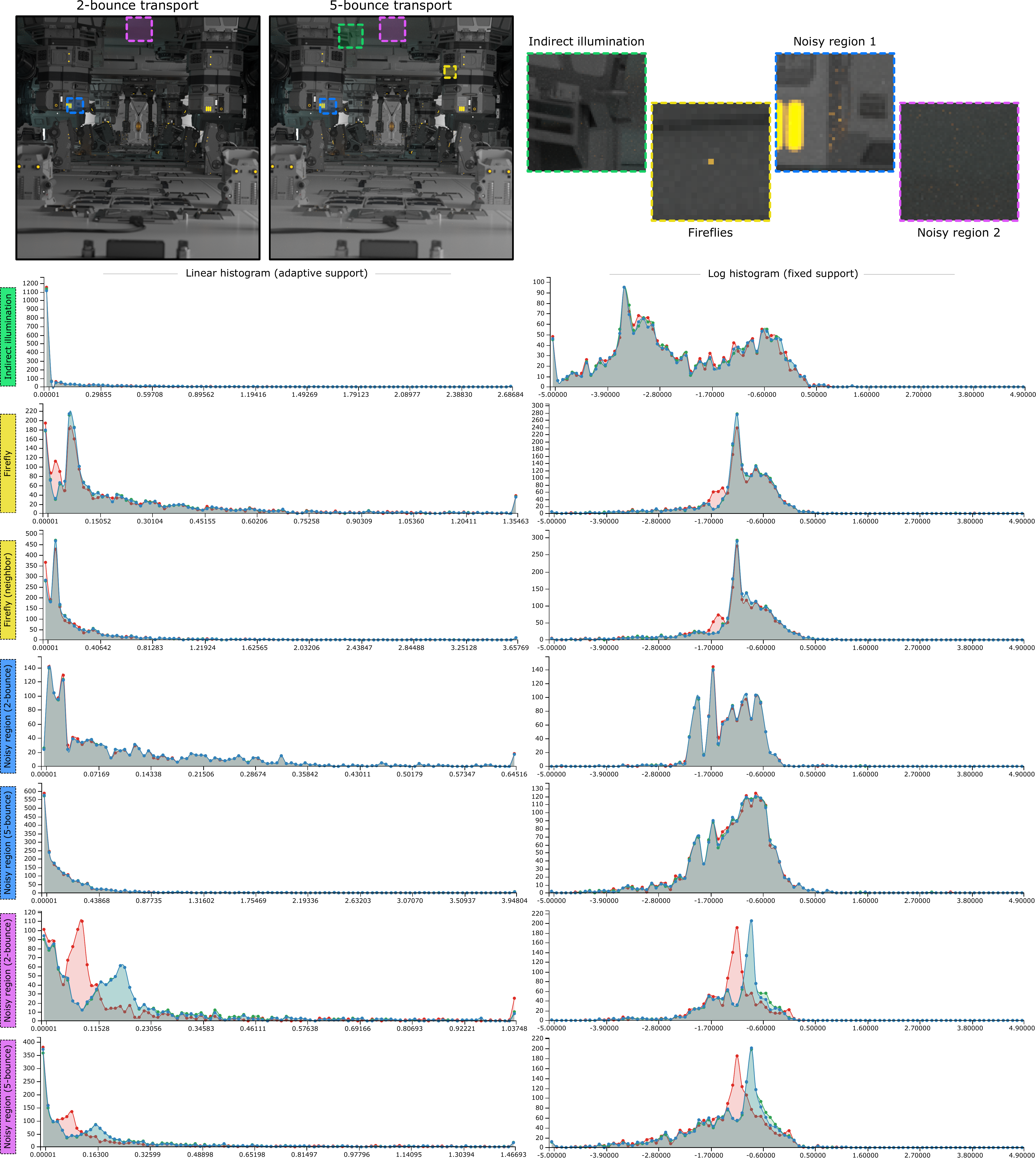}{t}{\studySize}{
Study for the \futuroScene{} scene, analyzed in \protect{\refSec{StudySceneD}}.
Note that in this figure, some of the histograms corresponding to the same regions (and color marks) are drawn from two different neighboring pixels \emph{(yellow)} or the two respective images \emph{(blue, magenta)}.
}

\mysection{Case Study}{CaseStudy}

Next, we present a detailed study of characteristic per-pixel radiance distributions.
We are primarily looking for recurring patterns in the distributions, as well as any other empirical properties that might guide the variance reduction model (\refEq{model}).

\Par{Instrumentation}
To facilitate the study, we developed a specialized web-based tool (which we plan to release with as a part of the publication) which allows for interaction with a specified rendered image, providing on-demand visualization and diagnostics of the characteristic distributions of the selected pixel, both in the linear and log-10 domains.
These data were extracted during rendering using 2k samples per pixel (SPP) of the next-event-estimation path tracer in PBRT3~\cite{Pharr2016}, and have the form of per-pixel RGB histograms with 100 bins for both the linear and log variants.

To overcome the challenge of setting the pixel-specific histogram support (\cf Delbracio~\citeetal{Delbracio2014}), we take the interval spanned by the first 50 collected samples and pad it by \percent{25} on both ends; for the log histogram we opted for the fixed \emph{absolute} support $[10^{-5},10^5]$ to complement the \emph{relative} information in their linear counterparts.
We count the null samples from occluded paths separately and do not include them in the visualized histograms, as they would cause discontinuous spikes without contributing much insight into the distributions' shapes.

\Par{Procedure}
We study the \emph{characteristic distributions} (CHD) corresponding to different objects, configurations and phenomena characteristic to every scene.
We examined each scene for interesting patterns, which we then present and discuss in the study but only if they are sufficiently abundant (\ie outlying distributions are not considered).
In the following subsections we study four main scenes selected from the standard distribution of PBRT3, depicted in \refFig{scenes_overview}.

%================================================================================================

\mysubsection{\bathroomScene{} Scene (\protect{\refFig{scenes_overview}a})}{StudySceneA}

We start with this simple interior scene illuminated by a single closed window; most of the illumination is therefore indirect.
There is a good variation of diffuse and glossy materials.
The transport in this scene is moderately difficult for a path tracer---after 2k SPP we get a reasonably converged image with only a few fireflies.

A selection of interesting characteristic distributions is depicted in \refFig{study_bathroom}.
We will now discuss these cases individually.
\begin{itemize}
	\item \textbf{Diffuse objects.} CHD tend to be quite symmetric in the log domain, close to Gaussian in fact.
	In the linear (radiance) domain, CHD exhibit long tails, strong skew, with mode typically lower than mean---resembling the Gamma-type distribution quite well.
	\item \textbf{Glossy objects.} Shapes surprisingly similar to diffuse objects, although in this case the different RGB channels' CHD are shifted \wrt each other (as a consequence of the underlying brown-wood texture).
	Interestingly, their spread differs across color channels, suggesting that their width is a function of the incoming radiance itself (as proposed in \refSec{PropertiesSampleDistributions}).
	\item \textbf{Noisy reflections.} All of the pixels in the reflection, in spite of their high estimate variance, share virtually the same CHD shape.
	This is in full agreement with the assumptions of Delbracio~\citeetal{Delbracio2014} regarding viable pixel similarity measures.
	\item \textbf{Direct illumination.} Very well resolved, low-noise CHD---in spite of their very significant spread and long tail.
	\item \textbf{Indirect illumination.} Noisy CHD with very long tails (high kurtosis).
	High number of null samples (around \percent{60}) and significant proportion of dark samples, showing inverted (negative) skew in the log domain (similar to the noisy reclection case).
	\item \textbf{Illumination gradients.} A visibly bi-modal CHD marking the transition from lit to shadowed regions.
	The relative proportions of the modes correspond to the strength of the penumbra.
\end{itemize}

%================================================================================================

\mysubsection{\whiteroomScene{} Scene (\protect{\refFig{scenes_overview}b})}{StudySceneB}
A moderately complex scene illuminated by three windows, sourcing a lot direct illumination but also having a high proportion of indirect illumination due to the highly reflective walls.
The illumination has a red color cast, both due to the outside source as well as the dark-red carpet.
It is predominantly populated with diffuse materials.
This scene is easily handled by a path tracer, due to the large open windows and low amount of geometric occlusion; this leads to low numbers of null samples as well.

\refFig{study_whiteroom} shows the characteristic distributions.
\begin{itemize}
	\item \textbf{Direct illumination.} White surfaces illuminated both directly and indirectly.
	This yields complex multi-modal CHD, with an added color cast for the aforementioned reasons.
	Very long tails are present, which is a common trait across the whole scene.
	\item \textbf{Diffuse objects.} Similar as the previous case, but with higher proportion of indirect illumination.
	This causes less spiky CHD in log domain and significantly longer tails in the primal domain.
	Significant noise is present as well, indicating more complex path structure in this region.
	\item \textbf{Indirect illumination.} In spite of the fact that these objects are spread across large space, their CHD look strikingly alike, as they uniformly reflect the CHD across the room.
	We again observe very long tails in the CHD, leading to somewhat higher variance in these areas compared to their directly illuminated counterparts.
	\item \textbf{Colorful objects.} These CHD follow the same patterns as already observed, but exhibit strong color cast due to the underlying diffuse texture.
	Note also, that the color cast shifts the distributions in the log domain, it basically does not change their shape---this makes the case that the material albedo has no correlation with variance and therefore makes for a poor feature descriptor for denoising purposes.
	\item \textbf{The Fireplace.} This fireplace appears boring, but after closer examination in the log domain displays a wild CHD, which, suspiciously, looks like fire.
	We are currently trying to interpret this behavior.
\end{itemize}

%================================================================================================

\mysubsection{\nightroomScene{} Scene (\protect{\refFig{scenes_overview}c})}{StudySceneC}
The same scene as in \refSec{StudySceneB}, but illuminated by two interior lights instead.
Although mostly indirectly illuminated, the lamps' translucent shades act as secondary `virtual' sources of direct light which then shows in the CHD.
This is a moderately difficult scene for a path tracer, due to the small and enclosed sources, leading to increased variance especcialy in the darker regions---consequently the proportion of null samples is high, often exceeding \percent{50}.

\refFig{study_nightroom} shows the characteristic distributions.
\begin{itemize}
	\item \textbf{Direct illumination.}
	These areas have bi-modal CHD composed of strong direct peaks (with a color shoft corresponding to the nearby illuminant) and faint indirect sample groups again with long tails.
	Further from the sources, the proportion of indirect samples grows, as expected.
	\item \textbf{Direct shadows.} In the linear domain, the CHD are very similar to the previous case, but missing the direct components.
	The log domain shows stronger skew towards the dark end, however, due to the higher complexity of the path structure.
	Consequently, we observe higher estimator variance in these regions.
	\item \textbf{Multi-bounce transport.} The CHD contain complex multi-modal patterns, but their shape is remarkably stable across channels, albeit with a color cast caused by the combination of reddish illuminants and green diffuse color of the objects themselves.
	\item \textbf{Virtual sources.} These are translucent lamp shades, which retain much of the direct component (compare to the first, directly illuminated case), but in addition contain a very long faint tail caused by the multi-scattered diffuse transmission.
	In spite of this, the variance in these areas is low, as the path tracer eventually does find the source inside for a sufficient proportion of the paths.
\end{itemize}

%================================================================================================

\mysubsection{\futuroScene{} Scene (\protect{\refFig{scenes_overview}d})}{StudySceneD}
A geometrically complex scene with neutrally colored mixture materials.
Illuminated mostly indirectly with area sources ranging from large background white panels down to small yellow sources.
Interestingly, even though this scene is geometrically very symmetric, the CHD are not in all cases, since the illumination differs for each half somewhat.
Though being a moderately difficult scene for a path tracer (with low numbers of null samples, typically under \percent{20}), the small yellow sources are difficult to find, creating patches of high variance around the image.

Due to the geometric complexity of this scene, we primarily use it for a special purpose: to study CHD with respect to statistical outliers (`fireflies') and with respect to different order of light transport (\ie how many bounces of indirect illumination are simulated by the path tracer).

\refFig{study_futuro} shows the characteristic distributions.
\begin{itemize}
	\item \textbf{Indirect illumination.} The complexity of this scene is reflected in its CHD as well.
	In this example, the linear histogram has such a wide support that most of its mass vanishes in a single spike; the log version however reveals a complicated multi-modal structure (although only a few modes are dominant, leaving hope that such shapes can still be learned).
	\item \textbf{Fireflies.} Here we compare two pixels: a bright firefly (RGB irradiance $[4.56, 2.03, 0.22]$) and its immediate neighbor on top (RGB irradiance $[0.25, 0.26, 0.26]$).
	In spite of this stark difference in value, their distributions are almost identical (which is better visible in the log histograms, as the linear variants differ in support).
	This is a behavior that we observe universally in all scenes, supporting the thesis of Delbracio~\citeetal{Delbracio2014} and justifying the reweighing strategies of Jung~\citeetal{Jung2015} and Zirr~\citeetal{Zirr2018}.
	One of the aims of our model is to learn such reweighing as well, but from data alone.
	\item \textbf{Different transport path lengths.} Here we compare the scene under identical configurations, with only one change: using two and five bounces of indirect transport respectively.
	As expected, that influences the amount of transported energy, making the order-2 image visibly darker than the order-5 one.\\
	The additional insight of our study is that the different path lengths also influence the shape of CHD: in accordance with the previous finding that higher orders of indirect illumination result in more complex path structures, indeed the resulting CHD for the order-5 images are more spread, with wider support and longer, noisy tails.
	The natural result is that the order-5 image ends up having higher variance, despite taking roughly twice longer to render.
	This finding agrees with theory (such as \refEq{VarE}), and points to the importance of discussing the complexity of the simulated transport in any MC variance reduction work.
\end{itemize}

%================================================================================================

\mysubsection{Findings of the Study}{StudyFindings}

We finish the study by drawing several general conclusions about the observed distributions.

Most importantly, \textbf{recurring patterns do exist in characteristic sample distributions}; despite being complex, CHD are certainly not random.
Moreover, CHD are not even discontinuous between neighboring pixels---if a radiance discontinuity lies in a given pixel, its CHD will end up a blend between the two regions of the function separated by that discontinuity (as indirectly implied in the work of Delbracio~\citeetal{Delbracio2014}).

The most frequent elementary distribution we have encountered across the board has a Gamma-like shape: a rapid onset with a long, slowly diminishing tail.
In the log domain, that shape appears more balanced and in fact not unlike a Gaussian.
As already hinted, this suggests a multiplicative nature of relations between samples.
In agreement with previous work~\cite{DeCoro2010, Jung2015, Zirr2018}, the long tails are the main culprit behind statistical fireflies and the generally difficult variance distributions so common in rendering.

The complexity of CHD is tightly linked with the scene complexity.
In more complex scenes, we observe multi-modal CHD, where each mode intuitively corresponds to a certain salient region of the path space.
The fact that light transport is a global problem with recursive structure means that the same patterns tend to occur all around the scene, repeatedly warped according to the respective geometric relations.
For that very reason, the path length influences the CHD complexity as well, with longer paths leading to distributions with larger support, longer tails and higher skew---and in result, higher residual variance.

The second most important finding, and the one with direct impact on the proposed variance-reduction model, is that \textbf{the shapes of the characteristic distributions are scene-specific}: they depend on the scene configuration holistically, rather than on a few key components or objects.
This suggests a clear design path: a variance reduction model aiming to take advantage of the CHD properties needs to be trained for each scene specifically (or even better, be able to classify the scene in a top-down manner and be conditioned by that classification).
In the following section we present a prototype of such a model, based on fully connected neural nets.

%================================================================================================
%================================================================================================

\myfigureDouble{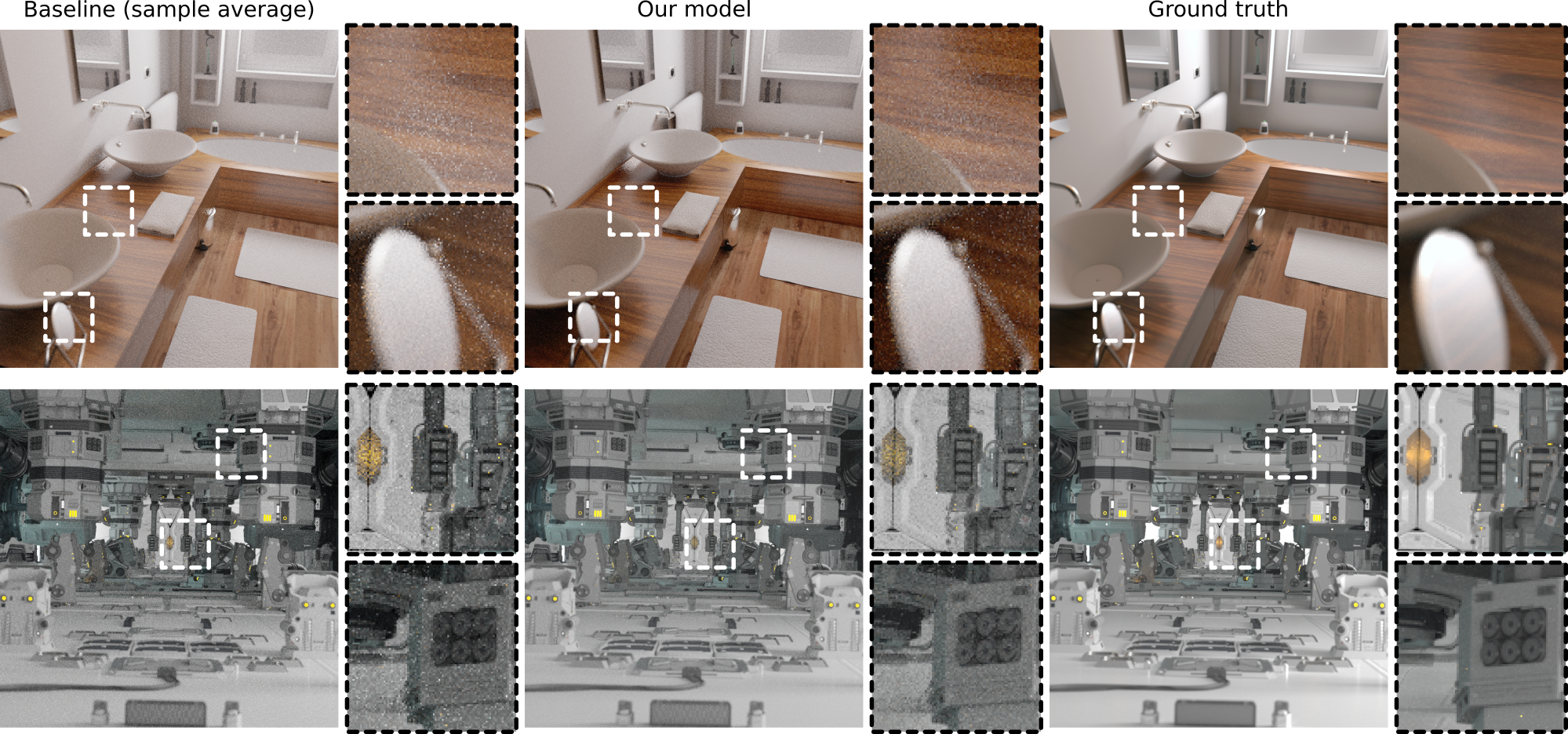}{
Variance reduction results for the \bathroomScene{} and \futuroScene{} scenes.
We compare the baseline estimate from 32 SPP and the result of our model for the same sample set, against the ground-truth images rendered with 2k SPP.
}

\mysection{Variance Reduction Model}{VarianceReductionModel}

The approach we propose---complementary to the existing lateral denoising approaches (\refSec{Background})---is to improve the estimates for $\myE_i$ (\refEq{Ei}) by training an \emph{a-posteriori} model $\model_\Theta$ defined by a set of parameters $\Theta$, according to the following criterion:
\begin{gather}
	\argmin_\Theta ~ \sum_i~\loss\left( \model_\Theta\left[ \{\est{\myL}(X)\}_i \right], \myE_i \right),
\label{eq:model}
\end{gather}
where $\loss$ is a suitable loss function, in our case a squared error between the model's response and the ground-truth irradiance for each pixel.
The assumption embedded in this strategy is that the per-pixel sample sets $\{\est{\myL}(X)\}_i$ generated by laws of light transport and the MC process that simulates them (path tracing, in our case) contain \emph{systematic patterns} that the model $\model$ can learn and exploit---and indeed, our study in \refSec{CaseStudy} has identified such patterns.
A reasonable expectation from such a model is that, in the absence of any identifiable patterns (or when they are too complex), it should gracefully reduce to the baseline solution, \ie the sample average as per \refEq{estEi}.
In its core, this is a Bayesian approach: start with a prior knowledge about the involved distributions and incorporate new data to create a posterior estimate.

\Par{Model implementation}
Our prototype model is a multi-layer perceptron network with a width-32 input layer (for the radiance samples $\est{\myL}$), three 32-neuron densely connected layers, and a single output unit representing the estimate.
In accordance with the findings of our study (\refSec{CaseStudy}) we trained a separate network for each tested scene.

The training process works with the sample data obtained from rendering the ground truth images using 2k samples per pixel.
For each training run, we uniformly selected 6400 pixels and divided their radiance samples into batches of 32 samples each; then separated out \percent{10} of the batches as validation data.
Using 30k training iterations with the learning rate of 0.005, the training took 20 minutes in Tensorflow on NVIDIA RTX 2080 Ti GPU.
The runtime of this model on an \resolutionX{800}{800} RGB image with \resolutionX{3}{32} SPP is under 1 minute in the unoptimized development mode.

\myfigure{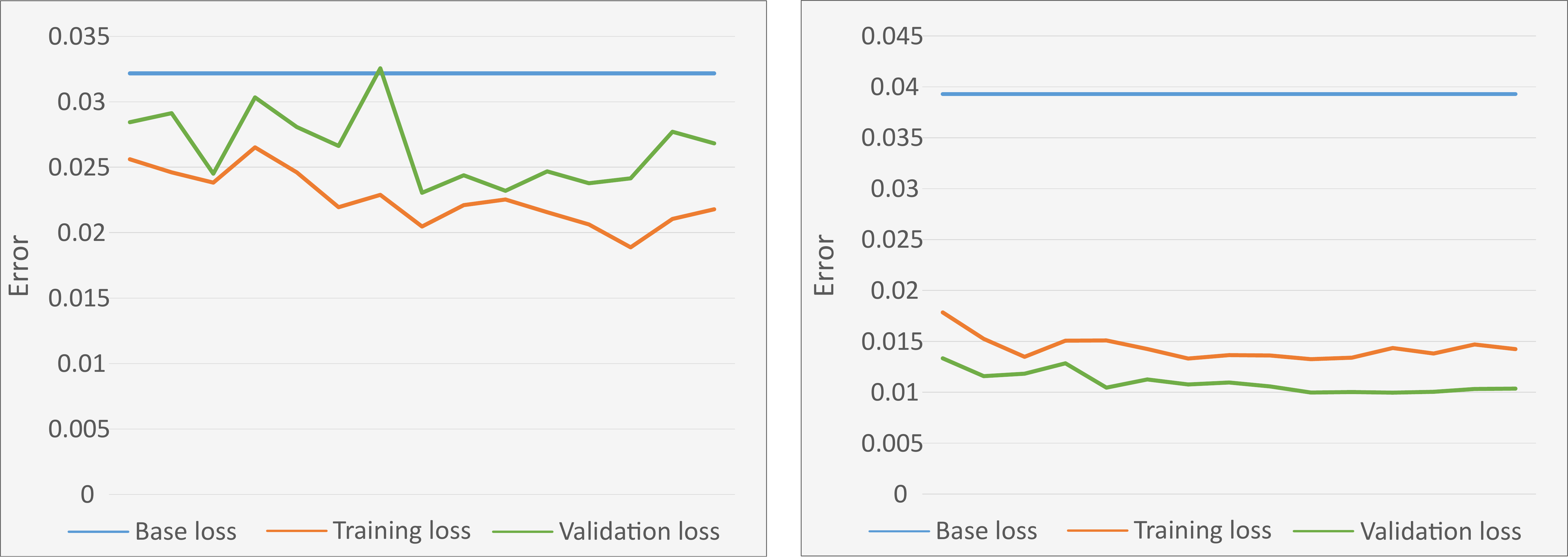}{
The loss function values obtained during the 30k training iterations for the \bathroomScene{} scene \emph{(left)} and the \futuroScene{} scene \emph{(right)}.
The loss for the training and validation data is plotted separately, against the baseline error.
}

\Par{Results}
The ability of our model to improve upon the baseline estimates is visually demonstrated in \refFig{denoising_results} for the \bathroomScene{} and \futuroScene{} scenes.
These results are promising, as our model yields smoother images even without exchaning any information between neighboring pixels.
In particular, a majority of the fireflies in the baseline is suppressed, a behavior similar to the results obtained by Zirr~\citeetal{Zirr2018}.
This has the benefit of regularizing the image to yield a more balanced variance distribution, which means that lateral denoising approaches would have an easier time using our model's output as the starting point.

To evaluate our model numerically, we first examined the quality of the results during training; this is shown in \refFig{plots_training_loss} for the two result scenes.
We see a reasonably stable behavior for both scenes, although surprisingly the model does not improve markedly after the initial iteration batch.
This indicates a space for future optimization of the training time, so that potentially the model could learn on the fly---that is, during the rendering process itself, to adapt to the current scene and rendering configuration.

The numerical counterpart to the evaluation in \refFig{denoising_results} is shown in \refFig{plots_numerical_eval}.
Here we compare the baseline results (sample averages) and our model to the ground-truth solution.
Although the majority of the estimates is improved by our model, some exceptions remain.
One possible way to address these would be to train the network to output a \emph{certainty} of each individual estimate---this is feasible as we know the deviations of the network's estimates from ground truth during training.
The certainty could then serve as a mixing weight for the model's output in combination with the baseline as fallback.

\myfigureCustom{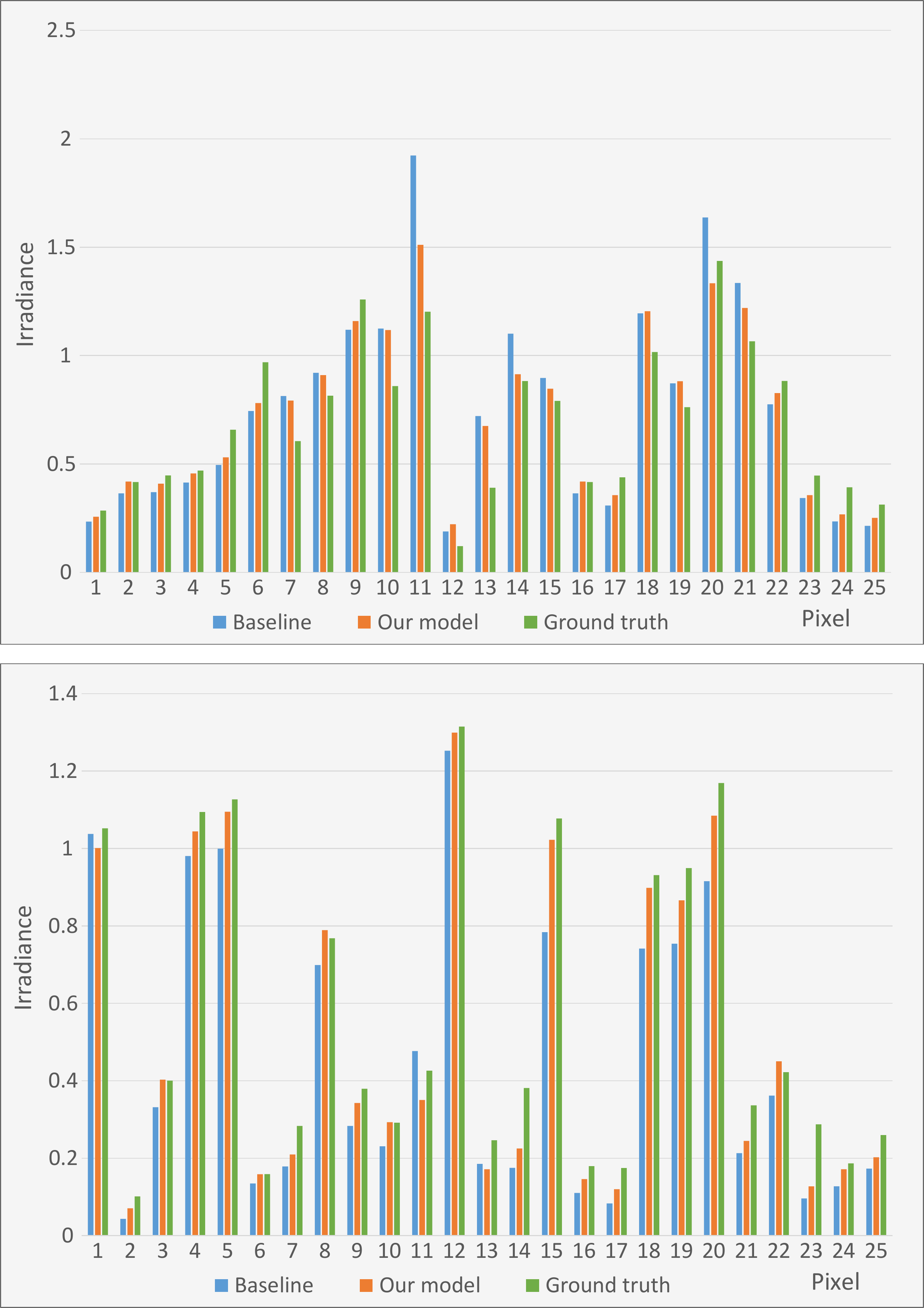}{tb}{1}{
Numerical evaluation for the \bathroomScene{} scene \emph{(top)} and the \futuroScene{} scene \emph{(bottom)}, shown for 25 randomly selected pixels from the respective images.
}

%================================================================================================
%================================================================================================

\mysection{Discussion}{Discussion}

\Par{Study extensions}
Our study identifies a number of properties regarding the characteristic sample distributions, as well as putting into question some of the commonly used assumptions about the MC variance reduction problem.
The aim of this study was to gain informed understanding about the problem, as well as inciting an open discussion about these issues.

A future prospect in improving the study involves both its breadth and depth.
More phenomena should be investigated---such as many-light configurations, caustics and participating media, to name a few---as well as more advanced integrators.
In particular, we wish to take a closer look at path guiding methods, as these are becoming state of the art in production and have to potential to subsume the majority of path-based methods.
More advanced visualization strategies might also be explored to extract further insights about the rendering process.

\Par{Model improvements}
Based on the obtained results, the prototype implementation of our proposed model already shows promise.
For the future we envision several directions in which we plan to improve upon it.

First, several quantitative improvements can be investigated.
The training and estimation timings are currently obtained from an unoptimized development implementation; we believe speedups of at least an order of magnitude are feasible.
Optimizing the network parameters would orthogonally improve the accuracy of the current estimates.

We also plan to investigate different network architectures, in particular recurrent NNs which have the potential of incorporating additional samples into the estimate as they are evaluated by the integrator; that would allow to run the model in a progressive rendering setting, with interactive refinements of the rendered image.

While naively extending the network to include neighboring samples would not be difficult, our aim was to show the possibility of reducing variance without this.
A compromise solution could operate in the spirit of Delbracio~\citeetal{Delbracio2014} and Boughida~\citeetal{Boughida2017}: the individual pixel estimators could share only \emph{statistical information} about the characteristic distributions.
Furthermore, a learning-based approach has the promise of incorporating high-level information about structures probabilistically detected in the image, such as edges, illumination gradients and even textures.

\Par{A-priori versus a-posteriori methods}
The presented approach tackles the problem of MC variance reduction in the \emph{a-posteriori} manner, trying to make sense of the collected estimator samples without making overly strong assumption about their distribution.
Due to the fact that we have the full information about the rendering process, bottom-up \emph{a-priori} approaches might seem favorable at first.
However, since we cannot theoretically reason about some of the parameters of the rendering process (particularly the scene), these approaches share the general `chicken-and-egg' problem of MC methods: the optimal solution to the problem involves knowing the solution upfront (another example is importance sampling).

In MC variance reduction, the problem of noisy pixel estimates is frequently deferred to the problem of noisy \emph{meta-estimates} of higher-order statistical quantities, or to idealized assumptions about their behavior.
We therefore argue that a certain amount of `data-drivedness' is necessary to break this loop and inform the denoiser about the properties of the rendered scene.
In future, we hope for the bottom-up methods (\eg \cite{Jung2015, Zirr2018}) to be unified with the learning-based approaches such as ours; a framework with the potential for this is arguably still yet to be found, but we believe that it should have Bayesian nature.

%================================================================================================
%================================================================================================

\mysection{Conclusion}{Conclusion}

Our approach is simple enough to be explained in one sentence: \textbf{we examine and learn the shapes of statistical sample distributions in Monte Carlo rendering, and apply that prior knowledge to produce improved estimates from finite sample sets}.

The case study we conducted confirms the expectation that per-pixel sample distributions typically have complex shapes that could hardly be represented by standard statistical distributions.
At the same time, we have identified recurring patterns that can be exploited by a learning-based model, along with the key finding that the characteristic per-pixel distributions holistically depend on the scene and the particular integrator used for the rendering.

The variance reduction model we present shows promising results, in spite of the fact that it has no information about the surrounding pixels.
Our plan is to continue our investigation and develop a truly robust model that is scene-sensitive, adapts to the particular rendering configuration, and runs progressively in lockstep with the rendering process.

\bibliographystyle{eg-alpha-doi}  
\bibliography{ms}

\newcommand{\etalchar}[1]{$^{#1}$}
\begin{thebibliography}{\uppercase{VRM{\etalchar{*}}18}}

\bibitem[BB17]{Boughida2017}
\textsc{Boughida M., Boubekeur T.}:
\newblock Bayesian collaborative denoising for {M}onte {C}arlo rendering.
\newblock \emph{Computer Graphics Forum (Proc. EGSR 2017) 36}, 4 (2017),
  137--153.

\bibitem[BCM05]{Buades2005}
\textsc{Buades A., Coll B., Morel J.~M.}:
\newblock A review of image denoising algorithms, with a new one.
\newblock \emph{SIAM Multiscale Model. Simul. 4}, 2 (2005), 490--530.

\bibitem[BRM{\etalchar{*}}16]{Bitterli2016}
\textsc{Bitterli B., Rousselle F., Moon B., Iglesias-Guiti\'{a}n J.~A., Adler
  D., Mitchell K., Jarosz W., Nov\'{a}k J.}:
\newblock Nonlinearly weighted first-order regression for denoising {M}onte
  {C}arlo renderings.
\newblock \emph{Computer Graphics Forum (Proc. EGSR) 35}, 4 (2016), 107--117.

\bibitem[BVM{\etalchar{*}}17]{Bako2017}
\textsc{Bako S., Vogels T., McWilliams B., Meyer M., Nov\'{a}k J., Harvill A.,
  Sen P., DeRose T., Rousselle F.}:
\newblock Kernel-predicting convolutional networks for denoising {M}onte
  {C}arlo renderings.
\newblock \emph{ACM Transactions on Graphics (Proc. of SIGGRAPH) 36}, 4 (2017).

\bibitem[CJ16]{Christensen2016}
\textsc{Christensen P.~H., Jarosz W.}:
\newblock The path to path-traced movies.
\newblock \emph{Found. Trends. Comput. Graph. Vis. 10}, 2 (2016), 103--175.

\bibitem[CKS{\etalchar{*}}17]{Chaitanya2017}
\textsc{Chaitanya C. R.~A., Kaplanyan A.~S., Schied C., Salvi M., Lefohn A.,
  Nowrouzezahrai D., Aila T.}:
\newblock Interactive reconstruction of {M}onte {C}arlo image sequences using a
  recurrent denoising autoencoder.
\newblock \emph{ACM Trans. Graph. 36}, 4 (2017), 98:1--98:12.

\bibitem[DMB{\etalchar{*}}14]{Delbracio2014}
\textsc{Delbracio M., Mus{\'e} P., Buades A., Chauvier J., Phelps N., Morel
  J.-M.}:
\newblock Boosting {M}onte {C}arlo rendering by ray histogram fusion.
\newblock \emph{ACM Trans. Graph. 33} (2014), 8.

\bibitem[DWR10]{DeCoro2010}
\textsc{DeCoro C., Weyrich T., Rusinkiewicz S.}:
\newblock Density-based outlier rejection in {M}onte {C}arlo rendering.
\newblock \emph{Computer Graphics Forum (Proc. Pacific Graphics) 29} (2010), 7.

\bibitem[GBBE18]{Guo2018}
\textsc{Guo J., Bauszat P., Bikker J., Eisemann E.}:
\newblock Primary sample space path guiding.
\newblock In \emph{Proc. of Eurographics Symposium on Rendering -- EI\&I}
  (2018), pp.~73--82.

\bibitem[GKH{\etalchar{*}}13]{Georgiev2013}
\textsc{Georgiev I., K\v{r}iv{\'a}nek J., Hachisuka T., Nowrouzezahrai D.,
  Jarosz W.}:
\newblock Joint importance sampling of low-order volumetric scattering.
\newblock \emph{ACM Transactions on Graphics 32} (2013), 6.

\bibitem[Hoo08]{Hoogenboom2008}
\textsc{Hoogenboom J.~E.}:
\newblock Zero-variance {M}onte {C}arlo schemes revisited.
\newblock \emph{Nuclear Science and Engineering 160}, 1 (2008), 1--22.

\bibitem[JMD15]{Jung2015}
\textsc{Jung J.~W., Meyer G., DeLong R.}:
\newblock Robust statistical pixel estimation.
\newblock \emph{Computer Graphics Forum 34}, 2 (2015), 585--596.

\bibitem[Kaj86]{Kajiya1986}
\textsc{Kajiya J.~T.}:
\newblock The rendering equation.
\newblock \emph{SIGGRAPH Comput. Graph. 20} (1986), 143--50.

\bibitem[KBS15]{Kalantari2015}
\textsc{Kalantari N.~K., Bako S., Sen P.}:
\newblock A machine learning approach for filtering {M}onte {C}arlo noise.
\newblock \emph{ACM Trans. Graph. 34}, 4 (2015).

\bibitem[KS13]{Kalantari2013}
\textsc{Kalantari N.~K., Sen P.}:
\newblock Removing the noise in {M}onte {C}arlo rendering with general image
  denoising algorithms.
\newblock \emph{Computer Graphics Forum (Proc. Eurographics) 32} (2013).

\bibitem[KW08]{Kalos2008}
\textsc{Kalos M.~H., Whitlock P.~A.}:
\newblock \emph{{M}onte {C}arlo methods}.
\newblock Wiley, 2008.

\bibitem[LMH{\etalchar{*}}18]{Lehtinen2018}
\textsc{Lehtinen J., Munkberg J., Hasselgren J., Laine S., Karras T., Aittala
  M., Aila T.}:
\newblock {N}oise2{N}oise: learning image restoration without clean data.
\newblock In \emph{Proc. of ICML} (2018), pp.~2965--2974.

\bibitem[MGN17]{Mueller2017}
\textsc{M\"{u}ller T., Gross M., Nov\'{a}k J.}:
\newblock Practical path guiding for efficient light-transport simulation.
\newblock \emph{Computer Graphics Forum 36}, 4 (2017), 91--100.

\bibitem[PJH16]{Pharr2016}
\textsc{Pharr M., Jakob W., Humphreys G.}:
\newblock \emph{Physically based rendering: from theory to implementation},
  3rd~ed.
\newblock Morgan Kaufmann, 2016.

\bibitem[RMZ13]{Rousselle2013}
\textsc{Rousselle F., Manzi M., Zwicker M.}:
\newblock Robust denoising using feature and color information.
\newblock \emph{Computer Graphics Forum} (2013).

\bibitem[SKW{\etalchar{*}}17]{Schied2017}
\textsc{Schied C., Kaplanyan A., Wyman C., Patney A., Chaitanya C. R.~A.,
  Burgess J., Liu S., Dachsbacher C., Lefohn A., Salvi M.}:
\newblock Spatiotemporal variance-guided filtering: real-time reconstruction
  for path-traced global illumination.
\newblock In \emph{Proc. of High Performance Graphics} (2017), pp.~2:1--2:12.

\bibitem[TM98]{Tomasi1998}
\textsc{Tomasi C., Manduchi R.}:
\newblock Bilateral filtering for gray and color images.
\newblock In \emph{Proc. of ICCV} (1998).

\bibitem[VKv{\etalchar{*}}14]{Vorba2014}
\textsc{Vorba J., Karl\'{\i}k O., \v{S}ik M., Ritschel T., K\v{r}iv\'{a}nek
  J.}:
\newblock On-line learning of parametric mixture models for light transport
  simulation.
\newblock \emph{ACM Trans. Graph. 33}, 4 (2014), 101:1--101:11.

\bibitem[VRM{\etalchar{*}}18]{Vogels2018}
\textsc{Vogels T., Rousselle F., McWilliams B., R\"othlin G., Harvill A., Adler
  D., Meyer M., Nov\'ak J.}:
\newblock Denoising with kernel prediction and asymmetric loss functions.
\newblock \emph{ACM Transactions on Graphics (Proc. SIGGRAPH 2018) 37}, 4
  (2018), 124:1--124:15.

\bibitem[XP05]{Xu2005}
\textsc{Xu R., Pattanaik S.~N.}:
\newblock A novel {M}onte {C}arlo noise reduction operator.
\newblock \emph{IEEE Computer Graphics and Applications 25}, 2 (2005), 31--35.

\bibitem[ZHD18]{Zirr2018}
\textsc{Zirr T., Hanika J., Dachsbacher C.}:
\newblock Re-weighting firefly samples for improved finite-sample {M}onte
  {C}arlo estimates.
\newblock \emph{Computer Graphics Forum (Proc. EGSR) 37}, 6 (2018), 410--421.

\bibitem[ZJL{\etalchar{*}}15]{Zwicker2015}
\textsc{Zwicker M., Jarosz W., Lehtinen J., Moon B., Ramamoorthi R., Rousselle
  F., Sen P., Soler C., Yoon S.-E.}:
\newblock Recent advances in adaptive sampling and reconstruction for {M}onte
  {C}arlo rendering.
\newblock \emph{Computer Graphics Forum (Proc. of EG State of the Art Reports)
  34}, 2 (2015), 667--681.

\end{thebibliography}

\end{document}